%% file: chapter.tex
\documentclass[graybox, envcountchap]{svmult}

\usepackage{mathptmx}        
\usepackage{amsmath}
\usepackage{amssymb}
\usepackage{color}
\usepackage{helvet}          
\usepackage{courier}         
\usepackage{dirtree}
\usepackage{journal_names}
\usepackage{makeidx}        
\usepackage{graphicx}        
\usepackage{subfig}

\usepackage{multicol}        
\usepackage[bottom]{footmisc}

\usepackage{hyperref}        
\hypersetup{colorlinks=true,urlcolor=blue}

\usepackage[misc]{ifsym}

\input{abb.tex}

\makeindex             

\begin{document}


\title{Tully-Fisher relation}
\author{Khaled Said}
\institute{Khaled Said (\Letter) \at School of Mathematics and Physics, University of Queensland, Brisbane, QLD 4072, Australia, \email{k.saidahmedsoliman@uq.edu.au}}
%
%
\maketitle

\abstract{The observed radial velocity of a galaxy consists of two main components: the recession velocity caused by the smooth Hubble expansion and the peculiar velocity resulting from the gravitational attraction of growing structures due to matter density fluctuations. To isolate the recession velocity component and calculate the Hubble constant, accurate measurements of true distances are needed. The Tully-Fisher relation is an empirical correlation between the luminosity and rotational velocity of spiral galaxies that serves as a distance indicator to measure distances independent of redshift. The Tully-Fisher relation has played an important role in Hubble constant measurements since its inception. This chapter delves into the significance of the Tully-Fisher relation in such measurements and explores its implications. We begin by discussing the definition and historical background of the Tully-Fisher relation. We also explore the observational evidence supporting this relation and discuss its advantages and limitations. The chapter then focuses on the methodology of using the Tully-Fisher relation for Hubble constant measurements. This includes detailed explanations of calibration techniques and biases. We emphasize the advantages of utilizing the Tully-Fisher relation, such as its ability to provide accurate distance measurements even at significant redshift where other methods may encounter challenges.}

\section{Introduction}
\subsection{Description of the Tully-Fisher relation}
\label{sec:1}

The Tully-Fisher (TF) relation is an empirical relation that correlates the intrinsic brightness of a spiral galaxy, measured by its total luminosity, and its dynamical properties, measured by its rotational velocity. In 1977, Brent Tully and Richard Fisher proposed the use of this relation as a distance indicator based on their observations of spiral galaxies in the Virgo cluster \cite{Tully-Fisher1977}. The main idea behind distance indicators is to use distance-independent observables to predict a distance-dependent parameter, which can subsequently be compared with the corresponding observable to derive a distance estimate. In the case of the Tully-Fisher relation, the rotational velocity of a spiral galaxy serves as the distance-independent observable that is used to predict the total absolute luminosity, which is distance-dependent. The total luminosity can then be compared with apparent magnitude to measure distance via the distance modulus. The Tully-Fisher relation has since been extensively studied and refined \cite{Silk1997,Avila-Reese1998,Kraan-Korteweg1986,O'Neil2000,McGaugh2000,McGaugh2005,Courteau2007,Said2015}. It has become a workhorse tool for measuring the distances and properties of galaxies, particularly in the context of large-scale structure surveys and cosmological studies \cite{Springob2007,Masters2008,Tully2009,Courtois2013,Hong2019,Courtois2023,Boubel2023}. 

\subsection{Historical Background}
The historical background of the Tully-Fisher relation can be traced back to the Great Debate in the 1920s, when Ernst \"{O}pik used an expression between the observed rotational velocity and the absolute magnitude to measure the true distance to Andromeda \cite{Opik1922}. This measurement helped to prove that Andromeda is an independent galaxy, not part of our own Milky Way Galaxy, as Shapley had thought. 

More than 50 years later, Balkowski et al used the Nan\c{c}ay radio telescope in France to measure the line widths of a sample of 13 irregular and spiral galaxies \cite{Balkowski1974}. They found a correlation between the line width and the luminosity but they did not apply this correlation as a distance indicator. This laid the foundation for the ground-breaking paper by Brent Tully and Richard Fisher in 1977 \cite{Tully-Fisher1977}. In their paper, Tully and Fisher used only inclined spiral galaxies and proposed the usage of the linear relation between \HI\ profile and absolute magnitude as a distance indicator. 

The publication of the Tully-Fisher relation in 1977 and the proposal to use it as a distance indicator was significant in many different ways. Firstly, it provided a robust new tool for measuring distance at redshifts that other methods such as Cepheid variable stars cannot reach. This opened up a whole range of large scale structure and cosmological studies. Notably, similar relations like the Fundamental Plane relation \cite{Djorgovski1987,Dressler1987} only emerged a decade later. Secondly, Tully and Fisher measured the Hubble constant $H_0$ to be 80 $\mathrm{km \ s^{-1} Mpc^{-1}}$ from the Virgo cluster and Ursa Major. This value was the first to deviate from the two mainstream values at that time, low value of $H_{0} = 50$ $\mathrm{km \ s^{-1} Mpc^{-1}}$ which was promoted by Sandage and Tammann \cite{Sandage1975} and a higher value of $H_{0} = 100$ $\mathrm{km \ s^{-1} Mpc^{-1}}$ claimed by de Vaucouleurs \cite{deVaucouleurs1979}.  This value is much closer to the value of Hubble constant that we know today.

\subsection{Tully-Fisher relation in cosmology}
The Tully-Fisher relation has been heavily used in measuring the Hubble constant $H_0$. Additionally, it has also been used as a valuable tool in many other cosmological studies\footnote{I do not intend to give a comprehensive review of all cosmological studies that used TF relation, but I will discuss a few examples before focusing on the role of TF relation in the $H_0$ tension which is the main focus of this chapter.}:
\begin{itemize}
    \item \textbf{Cosmography:} The Tully-Fisher relation uses the maximum rotational velocity of a galaxy to predict its luminosity. This can be used to estimate the distance to the galaxy, and to measure its peculiar velocity. The peculiar velocity is the deviation from the smooth Hubble flow and can be used to reconstruct 3D maps of the density and velocity fields of the universe. In 2014, a team led by Brent Tully used the Wiener filter reconstruction \cite{Zaroubi1995} to recover the 3D density and velocity maps. They used these maps to set the borders of our home supercluster which they called Laniakea \cite{Tully2014}. More recently, Alexandra Dupuy and H{\'e}l{\`e}ne Courtois used the full CosmicFlows-4 \cite{Tully2023} catalog\footnote{This catalog contains TF distances along with other distance indicators} to reconstruct the 3D density and velocity fields \cite{Courtois2023b} and then redefine the boundaries of the Laniakea supercluster along with other structures such as Perseus-Pisces and Shapley \cite{Dupuy2023}.  
    \item \textbf{Bulk Flow:} The bulk flow is the average peculiar velocity over a sphere of a given radius R. The theoretical expected value of this measurement relies on a model of a specific set of cosmological parameters, initial conditions and a chosen gravitational law \cite{Andersen2016}. In 2009, Watkins, Feldman and Hudson used the largest TF survey available at that time, the SFI++ survey \cite{Springob2007}, to measure the bulk flow of galaxies. They used a new method, the minimal variance estimator, and found a large bulk flow amplitude of $431\pm102$ km s$^{-1}$ within a scale of $50$ Mpc $h^{-1}$ \cite{Watkins2009}. The amplitude of this bulk flow is larger than expected from the $\Lambda$CDM model. More recently, several teams have used the state-of-the-art CosmicFlows-4 catalog \cite{Tully2023}, which contains more than 10,000 newly measured TF distances, to measure the bulk flow of galaxies. They have also found large bulk flows, arising on a much larger scale, that are higher than the expected value from the $\Lambda$CDM model \cite{Courtois2023b,Watkins2023,Whitford2023} 
    \item \textbf{Growth rate of structure:} Over the last three decades, many collaborations have started to test Einstein's general theory of relativity over a range of scales. One way to do that is by measuring the growth rate of cosmic structure, or $f\!\sigma_8$. This method is capable of measuring $f\!\sigma_8$ at low redshifts ($z < 0.1$), a range inaccessible to other  methods such as the redshift space distortion. The growth rate can be constrained by comparing the density field from redshift surveys and peculiar velocity fields from TF surveys. The growth rate can then be parameterized as a function of the mass density parameter $\Omega_m$ and the growth index $\gamma$, which is determined by the theory of gravity. In 2011, Marc Davis and others used the TF survey SFI++ to find a $f\!\sigma_8$ value of $0.31\pm0.06$ \cite{Davis2011}. Other teams also used the TF SFI++ catalog and found a slightly higher value of $0.401\pm0.024$ \cite{Carrick2015}. More recently, the CosmicFlows-4 catalog also was used to measure $f\!\sigma_8$ and found a similar value of $0.40\pm0.07$ \cite{Boubel2023}.  
    \item \textbf{Galaxy formation and evolution:} As an empirical relation between fundamental properties of galaxies, TF relation has been used to refine semi-analytical models and numerical simulations of galaxy formation and evolution \cite{Navarro2000,Vogelsberger2014,Ponomareva2018}.
    \item \textbf{Dark matter in galaxies:} The Tully-Fisher relation have also been used to probe the distribution and properties of dark matter in galaxies. Several studies put constraints on galaxy halo profiles by comparing the halo profile required by the Tully-Fisher relation to the density profile that is well-described by the Navarro-Frenk-White (NFW) profile \cite{Navarro1996,Mo2000,Seljak2002}.
\end{itemize}

\section{The Tully-Fisher relation}
\subsection{How does it work?}
The Tully-Fisher relation is defined and its usages in cosmology are discussed above. But how does it work in practice?

\begin{figure*}
	\includegraphics[width=\textwidth]{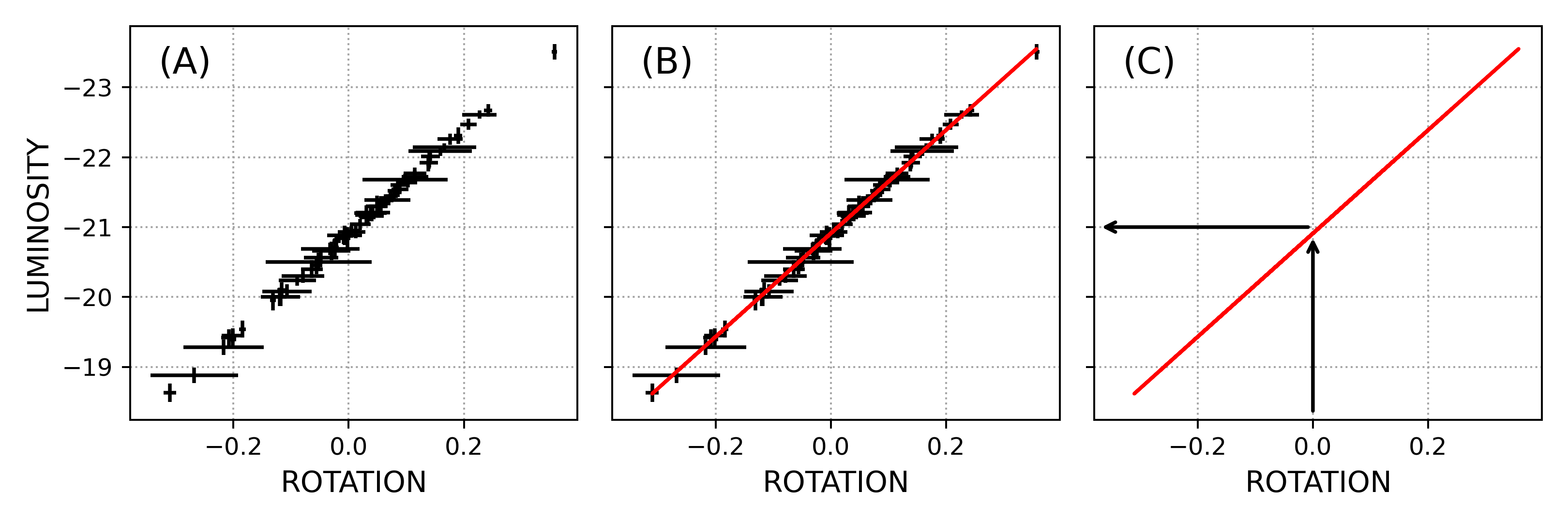}
    \caption{Illustrating how the Tully-Fisher relation combines photometric and spectroscopic data to enable the estimation of galaxy distance. In order to use the Tully-Fisher relation to measure distances, we need to build something called template relation. (A) shows a sample of galaxies with known rotation velocity, apparent magnitude, and distances which allows the determination of luminosity. (B) By fitting the TF relation to this sample, a template relation is derived. This establishes the correlation between the absolute magnitude and the rotational velocity of galaxies. (C) Applying the Tully-Fisher template to a galaxy of unknown distance, the absolute magnitude can be estimated based on its measured rotational velocity.}
    \label{TF_explained}
\end{figure*}

Building the Tully-Fisher relation requires two data sets: photometry and spectroscopy. Photometry provides a measurement of the galaxy's luminosity, which is the distance-dependent parameter in the TF relation. Spectroscopy provides a measurement of the galaxy's rotational velocity, which is the distance-independent parameter in the relation. The Tully-Fisher relation establishes the connection between the luminosity and rotational velocity of spiral galaxies. The TF relation is a secondary distance indicator and as the name suggests it requires calibration from a sample of galaxies with known distances usually coming from a primary distance indicator or a sample of cluster galaxies. Figure \ref{TF_explained}\textcolor{red}{A} represents the initial step in utilizing the TF relation for distance measurements. It shows a sample of galaxies with known rotational velocities and absolute magnitudes. The TF relation of the form:
\begin{eqnarray}
    M = a + b \log W
\end{eqnarray}
where $M$ is the absolute magnitude, $W$ represents the rotation, $a$ 
 and $b$ are the TF slope and intercept, respectively, can be fit to that sample. This relation, which we often call a template relation (Fig. \ref{TF_explained}\textcolor{red}{B}), can be used to predict the absolute magnitude of a galaxy with unknown distance given a measurement of its rotational velocity as shown in Fig. \ref{TF_explained}\textcolor{red}{C}. The predicted absolute magnitude can be compared to the galaxy's apparent magnitude to yield the distance via the distance modulus as:

 \begin{eqnarray}
     \mu = m - M(W) .
 \end{eqnarray}
    where $\mu$ is the distance modulus, $m$ is the apparent magnitude, and $M(W)$ is the predicted absolute magnitude of a galaxy from the TF relation given its rotational velocity

It is important to note that the derived template relation is not a universal relation. Instead, the data, parameters, and corrections used in the measurements of the TF distances should be consistent with those used in building the TF template relation. This has encouraged many teams to derive their own template relations instead of using an existing one. In the past few decades, different measurements, wavelengths, and methodologies have been used to do this. For example, shortly after the inception of the TF relation, several studies began to use near-infrared (NIR) bands because they suffer less than optical bands from dust extinction (both internal and external) and are more sensitive to stellar mass \cite{Aaronson1979,Masters2008,Said2015}. Other groups used isophotal magnitudes instead of total magnitudes to extend the Tully-Fisher studies into the Zone of Avoidance around the plane of the Milky Way \cite{Bouche2000,Said2016b}. 

In terms of spectroscopy, many parameters have been used in an attempt to minimise the intrinsic scatter of the TF relation. Many teams have used the 20\% \HI\ line width (the width measured at 20\% of the peak flux of the \HI\ profile) including the original TF relation \cite{Tully-Fisher1977,Pierce1988,Courtois2009}, while others suggested using the width measured at 50\% instead \cite{Giovanelli1997,Masters2006,Said2016a,Bell2023}. Furthermore, other teams suggest that the flat part of the optical rotation curve is the best proxy for the rotational velocity, and they have shown that it does indeed yield the tightest Tully-Fisher  relation \cite{Verheijen2001,Lelli2019}.

Different methodologies have been used to extract the best-fitting parameters for the Tully-Fisher  relation by applying different fitting procedures. In an attempt to overcome the Malmquist bias, Ren\'ee~C.~Kraan-Korteweg and others proposed the use of the inverse TF relation, which only takes into account the errors on the x-axis (rotational velocity) \cite{Kraan-Korteweg1986}. Since then, many teams have used both direct and inverse fitting procedures in their analyses \cite{Willick1995,Willick1996}. Other teams have used bivariate forms that take into account errors in both the x- and y-axes \cite{Giovanelli1997,Masters2008,Said2015,Bell2023}. Given the scatter in TF relation, a suggestion has been made to use a Bayesian mixture model, which is less biased by outliers \cite{Hogg2010}.

In summary, the Tully-Fisher relation has been observed to hold over a wide range of applications, including different parameters, methodologies, and wavelengths. This suggests that the TF relation is one of the most robust and powerful tools for measuring galaxy distances.

\subsection{Theoretical basis}
The physical origin of the Tully-Fisher  relation is still not fully understood, but it is certainly rooted in the physics of gravity and the dynamics of galactic rotation. The TF relation links the mass of a galaxy (characterized by the luminosity) to its dynamics (characterized by the rotational velocity). Specifically, it is based on the idea that the rotation velocity of a galaxy is related to the mass contained within a certain radius from the galaxy's center, known as the circular velocity.

Michael Strauss and Jeffrey Willick provided a widely accepted theoretical explanation\footnote{This was not the first such explanation, but it was particularly clear.} for the Tully-Fisher  relation in their 1995 review \cite{Strauss1995}.
They wrote that by equating the centrifugal force of an object moving in a circle of radius r to the gravitational attraction on the same object due to the mass inside a sphere of radius r, one can write

\begin{eqnarray}
    V_{\text{rot}}^2 \propto \frac{M}{r}
\end{eqnarray}
where $V_{\text{rot}}$ is the rotational velocity, $M$ is the mass within a sphere of radius $r$, and $r$ is the distance from the galaxy's center. Then, assuming a constant value for both the mass-to-light ratio (M/L) and mean surface brightness for spirals, one can obtain that

\begin{eqnarray}
    L \propto V_{\text{rot}}^4
\end{eqnarray}

However, most Tully-Fisher  studies have derived a power-law exponent that deviates from this theoretical explanation. This is usually attributed to the complication of the dark matter halo that surrounds a galaxy \cite{Mould2020}. Early near-infrared Tully-Fisher analyses showed that the complications of halo mass could be avoided, and indeed derived a power-law exponent that was closer to the theoretical expectation \cite{Aaronson1979}. However, their conclusion has been contradicted by many larger subsequent surveys \cite{Giovanelli1997,Masters2008}.

In 2000, Stacy McGaugh and his team investigated the deviation of dwarf faint galaxies from the linear Tully-Fisher  relation \cite{McGaugh2000}. They found that the linear TF relation could be restored if they used the sum of both stellar and gas mass components instead of luminosity. This relation takes the form

\begin{eqnarray}
    M_d \propto V_{\text{rot}}^4
\end{eqnarray}

where $M_d$ is the sum of stellar and gas masses,

\begin{eqnarray}
    M_d = M_{*} + M_{\text{gas}}.
\end{eqnarray}

Not only did they establish the baryonic TF relation as the fundamental relation, but they also found that the intrinsic scatter was smaller than the original TF relation. This result can be simply explained if we suppose that this estimate of the baryonic mass is a better proxy for the true total mass than the luminosity or the stellar mass alone. 

This result, at first glance, may seem to support the Modified Newtonian Dynamics (MOND: \cite{Milgrom1983}) as an alternative to dark matter. MOND is a theory of gravity that modifies Newtonian dynamics. In MOND, the baryonic mass is actually the total mass, which means that the baryonic TF relation is the fundamental TF relation. However, it is important to note that Modified Newtonian Dynamics (MOND) has its own critics \cite{vandenBosch2000,vandenBosch2001}. We refer the reader to the literature for more information about the debate between MOND and Cold Dark Matter.

One important theoretical development in recent years has been the use of hydrodynamic simulations to study the formation and evolution of galaxies in a cosmological context \cite{Vogelsberger2014,Schaye2015}. These simulations incorporate the effects of gravity, gas dynamics, star formation, and feedback, and have been shown to predict the Tully-Fisher relation in different environments and at different redshifts better than semi-analytical models \cite{Papastergis2016}.

\subsection{Advantages and limitations}
The Tully-Fisher relation has several advantages as a tool for measuring distances of galaxies. The first and main advantage is its simplicity and robustness. The TF relation provides a linear correlation between only two easily measurable galaxy properties, luminosity and rotational velocity. Several studies have shown that, unlike other distance indicators, the scatter of the TF relation is relatively insensitive to other galaxy properties \cite{Courteau1999,Courteau2007,Ouellette2017}. This makes the TF relation a fundamental, robust and reliable tool for measuring distances. 

Another advantage of the Tully-Fisher  relation is that it uses spiral galaxies, which are the most numerous type of galaxies at low redshifts ($z<0.07$). This makes it a convenient tool for measuring peculiar velocities at these redshifts, where the true distances can be determined. Beyond this redshift, it is more difficult to obtain accurate peculiar velocities as the errors in measuring these velocities from distance indicators scale with redshift. 

\begin{figure}
\sidecaption
\includegraphics[scale=.465]{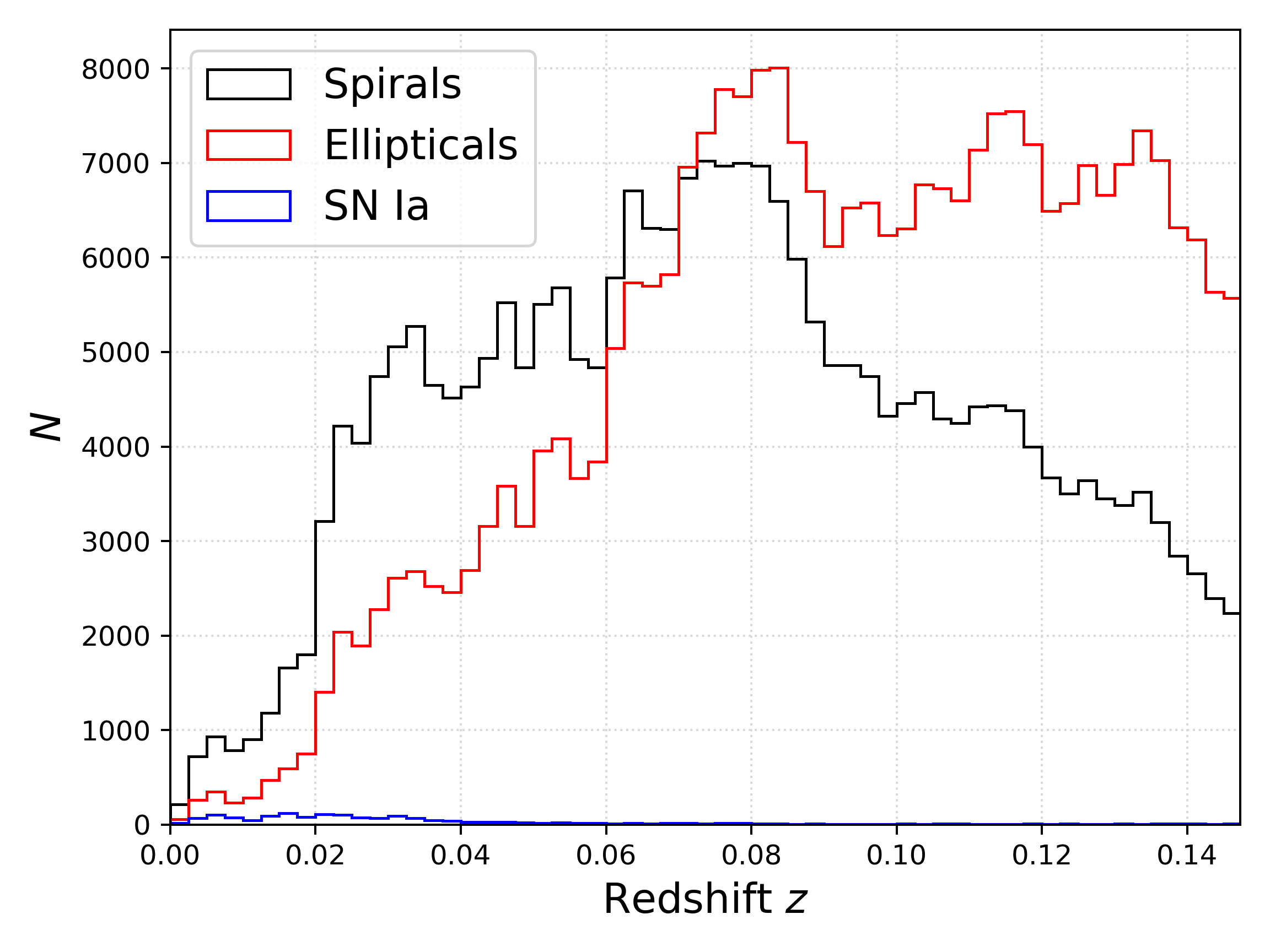}
\caption{The redshift distribution of spiral (black) and elliptical (red) galaxies in the SDSS survey, as well as the redshift distribution of the state-of-the-art SN Ia from Pantheon+ (blue), up to redshift $z<0.15$. The number of galaxies per 0.0025 redshift bin is shown.}
\label{s_e_sn} 
\end{figure}

In Figure \ref{s_e_sn}, I show the redshift distribution of spiral and elliptical galaxies from the SDSS survey DR14 \cite{Abolfathi2018}. I separated spirals from ellipticals using the galaxy zoo catalog \cite{Lintott2011}. I also over plot the redshift distribution of the state-of-the-art Pantheon+ sample \cite{Carr2022,Brout2022}. I used only the redshift below $z<0.15$, as this is the range used for Hubble constant ($H_0$) measurements \cite{Riess2016}. It is clear that spiral galaxies (Tully-Fisher galaxies) are the most numerous type at low redshift $(z< 0.07)$, with a mean redshift of $z=0.077$. This makes TF relation one of the most precise tools for measuring $H_0$ at this redshift.  

However, there are also some limitations to the Tully-Fisher  relation as a tool for distance measurements. One of the main limitations of the TF relation is its intrinsic scatter. This scatter can lead to errors of around $\sim$20\% in distance measurements. The baryonic Tully-Fisher relation is often shown to be a tighter relation than the TF relation, with a much smaller intrinsic scatter \cite{McGaugh2012,Lelli2016}. However, measuring the sum of the stellar and gas components is much more difficult than measuring just the luminosity of a galaxy. This adds a lot of observational uncertainties to the baryonic TF relation.

Another limitation of the Tully-Fisher  relation and a significant source of uncertainty is the inclination of galaxies. The inclination of a spiral galaxy affects both TF parameters. It affects the derived maximum rotational velocity during the process of correcting the observed line width for projection effects (i.e. to edge-on orientation). Additionally, to correct the magnitude for the internal extinction, one needs to use the inclination to account for the path-length of the light through the galaxy. Therefore, inclination errors lead to correlated errors in the two TF parameters that have the same sense (qualitatively) as errors in distance, so an error in inclination is (at least partly) degenerate with an error in distance. Most Tully-Fisher surveys select only edge-on galaxies by applying a cut to inclinations in an attempt to minimize the line width correction. Although this seems like the right thing to do, inaccurate inclinations will lead to a selection bias in the TF distance. This is because galaxies with inaccurate inclinations may be excluded or included in the survey, depending on whether their inclinations are overestimated or underestimated. Furthermore, edge-on galaxies are subject to higher internal extinction, which means that more light is blocked by dust and gas inside the galaxy. This results in a dimmer magnitude, and thus a larger correction is needed to correct for the extinction. Some studies have even suggested that when the inclination measurements have errors larger than 10 degrees, it may be more prudent to omit inclinations entirely rather than assuming them to be exact \cite{Obreschkow2013}.

Selection effects and Malmquist bias are other limitations of the TF relation. However, these limitations are not specific to the TF relation, as they affect all distance indicators.

\section{The Role of the Tully-Fisher relation in $H_0$ Measurements}
The methodology for using the Tully-Fisher relation in $H_0$ measurements typically involves the following steps. First, a sample of galaxies with well-measured rotational velocities and luminosities (for TFR) or masses (for the BTFR) is selected (see Fig. \ref{TF_explained}\textcolor{red}{A}). These galaxies are typically chosen to be representative of the population of galaxies being studied. Next, the Tully-Fisher relation is calibrated for this sample of galaxies. This involves measuring the slope, intercept, and scatter of the relation for the sample, and correcting for any systematic biases or selection effects (see Fig. \ref{TF_explained}\textcolor{red}{B}). Once the Tully-Fisher relation has been calibrated for the sample, it can be used to infer the distances to other galaxies with similar properties (see Fig. \ref{TF_explained}\textcolor{red}{C}). The final step is measuring $H_0$, which can be done by plotting the derived TF distances against redshifts.

This methodology has been used in a number of studies to estimate the value of the Hubble constant ($H_0$). In 1977, Tully and Fisher applied their newly discovered Tully-Fisher relation to derive distances to the Virgo and Ursa Major clusters. They measured a Hubble constant of $H_0 = 84$ km s$^{-1}$ Mpc$^{-1}$ for Virgo and $H_0 = 75$ km s$^{-1}$ Mpc$^{-1}$ for Ursa Major. They conclude a preliminary value for the Hubble constant of $H_0 = 80$ km s$^{-1}$ Mpc$^{-1}$. Although they called it preliminary and did not propagate their distance error to the Hubble constant value, we know today that it was the most accurate result at that time. 

In 1983, Visvanathan used the Tully-Fisher relation to measure distances to 52 cluster spiral galaxies. They found that the Hubble constant derived from nearby clusters varied from 58.5 to 83.5, but the range was much smaller for distant clusters, between 76.3 and 78.9 km s$^{-1}$ Mpc$^{-1}$. Visvanathan concluded that the best fit value for the Hubble constant from all clusters in their study was $H_0 = 74.4\pm11$ km s$^{-1}$ Mpc$^{-1}$ \cite{Visvanathan1983}. That result is very close to the value of $H_0$ we have today.

Shortly after, Sandage and Tammann used the Tully-Fisher relation in the infrared (IR) and blue bands to estimate distances to galaxies in the Virgo and Coma clusters, respectively. Their analysis gave a lower value of the Hubble constant $H_0 = 55\pm9$ km s$^{-1}$ Mpc$^{-1}$ \cite{Sandage1984}. This result aligned with their previous measurement of $H_0 = 50\pm7$ km s$^{-1}$ Mpc$^{-1}$ obtained through Type I supernovae \cite{Sandage1982}.

Aaronson led a team in 1986 to derive distances to 10 galaxy clusters using the infrared Tully-Fisher relation. They stated that the current evidence favors a large value for $H_0$ of 90 km s$^{-1}$ Mpc$^{-1}$ \cite{Aaronson1986}. In two subsequent studies, Bottinelli led a team to use the infrared Tully-Fisher relation to derive distances to galaxy clusters. They concluded that the $H_0$ lies between 70 and 75 km s$^{-1}$ Mpc$^{-1}$. During the same time, Ren\'ee~C.~Kraan-Korteweg and others derived distances to a larger sample of 82 cluster galaxies using both the infrared Tully-Fisher relation and the optical Tully-Fisher relation. They found that their data is consistent with a Hubble constant of $H_0 = 56.6 \pm 0.9$ km s$^{-1}$ Mpc$^{-1}$ \cite{Kraan-Korteweg1988}. However, this low value was contradicted by three studies. One study used the TF relation to measure distances to Virgo and Ursa Major clusters, and derived a value of $H_0 = 85 \pm 10$ km s$^{-1}$ Mpc$^{-1}$ \cite{Pierce1988}. The other two studies derived distances to the Coma cluster, and also found a higher value of $H_0 \approx 90$ km s$^{-1}$ Mpc$^{-1}$ \cite{Fukugita1991,Rood1993}. Bureau, Mould, and Staveley-Smith \cite{Bureau1996} used the Tully-Fisher relation to measure distances to galaxies in the Fornax cluster, and derived a value of $H_0 = 74 \pm 11$ km s$^{-1}$ Mpc$^{-1}$. 

This significant discrepancy in these early measurements can be attributed to various factors, including the challenge of sample incompleteness correction and the absence of reliable distances to nearby objects for use as a calibration sample.

The Hubble Space Telescope Key Project on the Extragalactic Distance Scale  used Cepheid distances to re-calibrate the Tully-Fisher relation. In a series of papers, they found results consistent with a value of $H_0$ of $71\pm4$ km s$^{-1}$ Mpc$^{-1}$ \cite{Madore1999,Sakai2000,Freedman2001}.

The Cosmicflows project combines distances measured through various methods, including the Tully-Fisher relation, surveys, and collaborations. This project has made significant contributions to measuring the Hubble constant, consistently obtaining values in the range of $74\pm4$ to $76\pm1$ km s$^{-1}$ Mpc$^{-1}$ over an extended period \cite{Tully2012,Sorce2012,Sorce2013,Neill2014,Sorce2014,Kourkchi2020,Kourkchi2022,Courtois2023b}

\begin{figure*}
	\includegraphics[width=\textwidth]{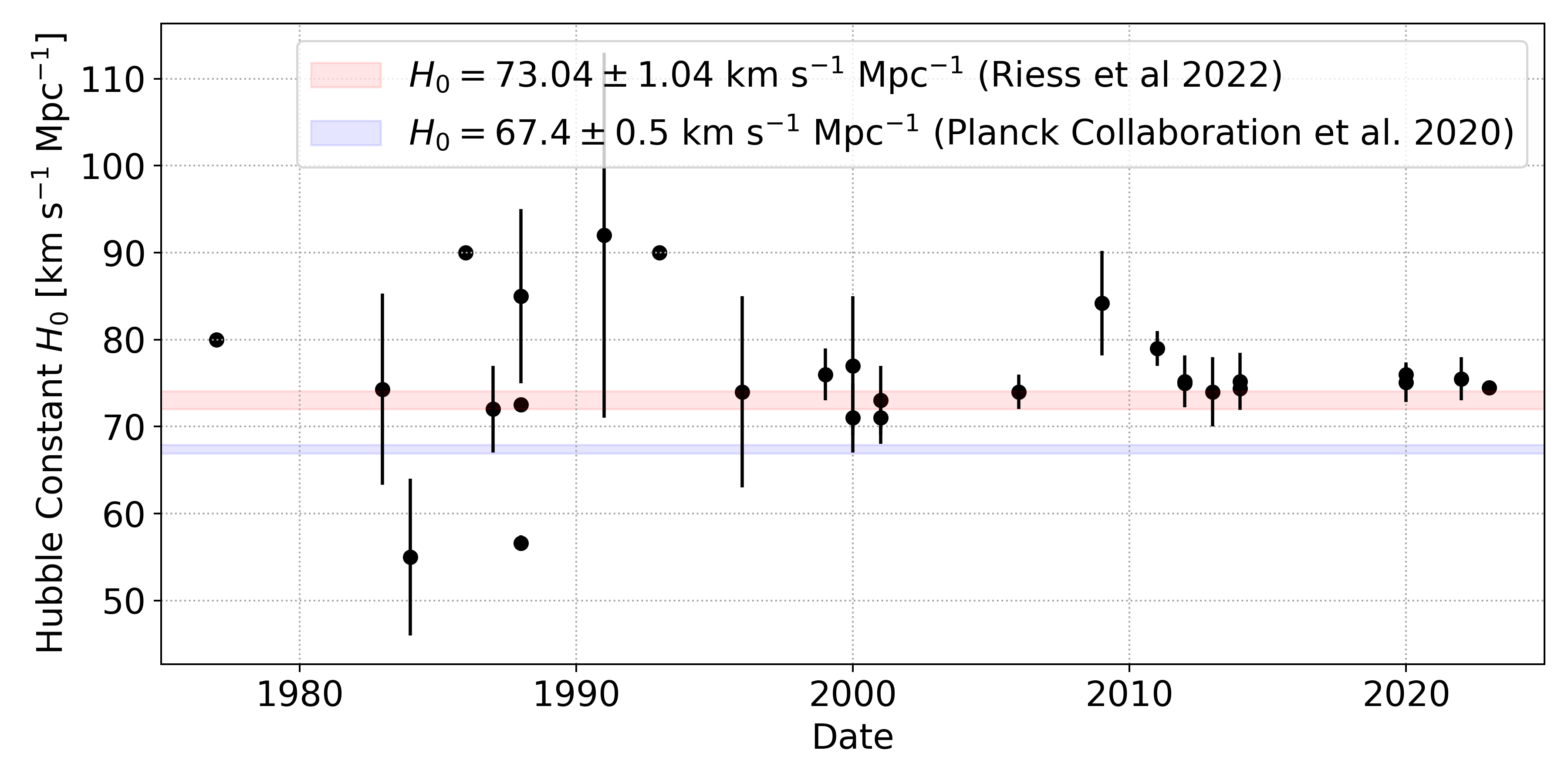}
    \caption{A sample of published Hubble constant ($H_0$) measurements using Tully-Fisher relation spanning from 1977 to 2023. Most of these measurements are obtained by different teams using different data given in table \ref{tab:1}. The red band highlights the most recent Hubble constant value and its associated uncertainty from the baseline results of the Cepheid-SN Ia sample by the SH0ES team \cite{Riess2022}. The blue band represents the Hubble constant value reported by Planck Collaboration 2020 \cite{PlanckCollaboration2020}. The majority of the data points align closely with the red band, particularly from 2000 onward. It is important to note that this sample of data is just a subset from the literature and additional studies may already exist beyond those represented here.}
    \label{H0_explained}
\end{figure*}

Figure \ref{H0_explained} shows a graphical representation of the published efforts to measure the Hubble constant ($H_0$) using the Tully-Fisher relation. The red band shows the the most recent result of Hubble constant and its associated errors measured using Type Ia supernovae by the SH0ES team in 2022 \cite{Riess2022}. Additionally, the blue band represent the Hubble constant value and its associated error from the CMB measured by the Planck Collaboration in 2020 \cite{PlanckCollaboration2020}. The figure shows a lot of scatter in the TF results prior to 2000 with some unrealistically optimistic uncertainties. However, after 2000, the results seem consistent and in more agreement with the value derived by the SH0ES team in 2022 \cite{Riess2022} than the CMB value by the Planck Collaboration in 2020 \cite{PlanckCollaboration2020}. Table \ref{H0_table} provides a summary of the results from these studies.

\begin{table}
\caption{Evolution of published Hubble constant values from the Tully-Fisher relation (1977-2023)}
\label{tab:1}       
\begin{tabular}{p{1.5cm}p{2cm}p{1.7cm}||p{1cm}p{1.5cm}p{2cm}p{1.3cm}}
\hline\noalign{\smallskip}
Date & $H_0$ & Ref. & & Date & $H_0$ & Ref.  \\
\noalign{\smallskip}\svhline\noalign{\smallskip}
  1977 & 80.0  & \cite{Tully-Fisher1977} & & 2001 & 71.0 $\pm$ 3.0 & \cite{Freedman2001}\\
  1983 & 74.3 $\pm$ 11.0 & \cite{Visvanathan1983} & & 2001 & 73.0 $\pm$ 4.0 & \cite{Watanabe2001}\\
  1984 & 55.0 $\pm$ 9.0 & \cite{Sandage1984} & & 2006 & 74.0 $\pm$ 2.0 & \cite{Masters2006}\\
  1986 & 90.0  & \cite{Aaronson1986} & & 2009 & 84.2 $\pm$ 6.0 & \cite{Russell2009}\\
  1987 & 72.0 $\pm$ 5.0 & \cite{Bottinelli1987} & & 2011 & 79.0 $\pm$ 2.0 & \cite{Hislop2011}\\
  1988 & 85.0 $\pm$ 10.0 & \cite{Pierce1988} & & 2012 & 75.0  & \cite{Tully2012}\\
  1988 & 72.5  & \cite{Bottinelli1988} & & 2012 & 75.2 $\pm$ 3.0 & \cite{Sorce2012}\\
  1988 & 56.6 $\pm$ 0.9 & \cite{Kraan-Korteweg1988} & & 2013 & 74.0 $\pm$ 4.0 & \cite{Sorce2013}\\
  1991 & 92.0 $\pm$ 21.0 & \cite{Fukugita1991} & & 2014 & 74.4 $\pm$ 1.0 & \cite{Neill2014}\\
  1993 & 90.0  & \cite{Rood1993} & & 2014 & 75.2 $\pm$ 3.3 & \cite{Sorce2014}\\
  1996 & 74.0 $\pm$ 11.0 & \cite{Bureau1996} & & 2020 & 75.1 $\pm$ 2.3 & \cite{Schombert2020}\\
  1999 & 76.0 $\pm$ 3.0 & \cite{Madore1999} & & 2020 & 76.0 $\pm$ 1.1 & \cite{Kourkchi2020}\\
  2000 & 77.0 $\pm$ 8.0 & \cite{Tully2000} & & 2022 & 75.5 $\pm$ 2.5 & \cite{Kourkchi2022}\\
 2000 & 71.0 $\pm$ 4.0 & \cite{Sakai2000} & & 2023 & 74.5 $\pm$ 0.1 & \cite{Courtois2023b}\\
\noalign{\smallskip}\hline\noalign{\smallskip}
\end{tabular}
\label{H0_table}
$^a$ This is just a subset of the published Hubble constant values using TF relation, and additional studies may already exist beyond those represented here.
\end{table}

In summary, the Tully-Fisher relation has been used for decades to estimate distances to galaxies and to measure the Hubble constant. Although Tully-Fisher measurements of $H_0$ have shown a lot of scatter prior to 2000, they have been consistent since then with a value of $H_0$ that agrees with the other low-redshift probes.

\section{Future Prospects}
The largest homogeneous Tully-Fisher  sample currently exists is the CosmicFlows-4 survey \cite{Kourkchi2020}. This survey includes 10,000 TF distances. Imminent surveys such as DESI \cite{Saulder2023} and WALLABY \cite{Courtois2023} will provide 50,000 and 200,000 TF distances, respectively, over the coming few years. With approximately 20\% TF distance errors for about 250,000 galaxies covering a significant portion of the sky up to redshift $z\approx0.1$, both random and systematic errors in $H_0$ from large-scale structure will be considerably smaller compared to the most optimistic uncertainties associated with other low-redshift probes.

\begin{acknowledgement}
KS acknowledges support from the Australian Government through the Australian Research Council’s Laureate Fellowship funding scheme (project FL180100168). I would like to express my gratitude to Tamara Davis and Matthew Colless for their valuable feedback on the initial draft of this review.
\end{acknowledgement}



\bibliographystyle{unsrt2authabbrvpp}
\bibliography{ref}


\end{document}

%% file: abb.tex
\def\la{\mathrel{\hbox{\rlap{\hbox{\lower4pt\hbox{$\sim$}}}\hbox{$<$}}}}
\def\ga{\mathrel{\hbox{\rlap{\hbox{\lower4pt\hbox{$\sim$}}}\hbox{$>$}}}}

\newcommand{\HI}{\mbox{\normalsize H\thinspace\footnotesize I}}

\def\apjl   {{ApJL }}
\def\apjs   {{ApJS }}